\documentclass[fleqn,twoside]{article}
\usepackage{espcrc2}
\usepackage{graphicx}
\usepackage{graphicx}
\usepackage[figuresright]{rotating}

\newlength{\bredde}
\def\slash#1{\settowidth{\bredde}{$#1$}\ifmmode\,\raisebox{.15ex}{/}
\hspace*{-\bredde} #1\else$\,\raisebox{.15ex}{/}\hspace*{-\bredde} #1$\fi}
\newcommand{\beq}{\begin{equation}}
\newcommand{\eeq}{\end{equation}}
\newcommand{\AmS}{{\protect\the\textfont2
  A\kern-.1667em\lower.5ex\hbox{M}\kern-.125emS}}

\title{\vspace{-4.0cm}
       \rightline{\normalsize CERN-TH/2001-289}
       \rightline{\normalsize October 2001}
       \vspace{2.5cm}
The Microscopic Dirac Operator Spectrum}

\author{P.H. Damgaard\address{Theory Division, CERN, CH-1211 Geneva 23, 
        Switzerland}%
        \thanks{On leave from the Niels Bohr Institute, Blegdamsvej 17, 
        DK-2100 Copenhagen, Denmark.}}
       
\begin{document}

\begin{abstract}
We review the exact results for microscopic Dirac operator spectra
based on either Random Matrix Theory, or, equivalently, chiral Lagrangians.
Implications for lattice calculations are discussed.
\vspace{1pc}
\end{abstract}

\maketitle

\section{Introduction}

Exact statements can be made about the distributions of the smallest 
Dirac operator eigenvalues in
gauge theories with spontaneous chiral symmetry breaking.
This extends previous exact results
for fermion zero modes in topologically non-trivial gauge field backgrounds
to an infinite series of non-zero modes.
In this short review we
focus exclusively on these exact results for the smallest
Dirac operator eigenvalues in four dimensions, and discuss why they are so
important for lattice gauge theory. 

The one single assumption that has to be made is that chiral
symmetry is spontaneously broken. Actually, this statement needs
a small qualification because the group of chiral symmetries
depends on the number of fermions, and
on the representation carried by them. 
We can be completely general and consider $N_f$ (Dirac) fermions in an
arbitrary representation $r$ of the gauge group ${\cal G}$ with
vectorlike couplings.
As one could suspect from Wigner's classification of representations,
there are just three symmetry breaking classes to consider.
It all depends on whether the representation is complex, real or
pseudo-real. The case of complex 
representations is not only the
physically most relevant (QCD belongs to this class), it is also
the most simple. In the two other cases the initial symmetry is larger
than one would naively have expected, namely SU(2$N_f$) rather than
just SU$_L(N_f)\times$SU$_R(N_f)$. 
The expected patterns of chiral symmetry breaking are as
follows, depending on the representation $r$:

\begin{itemize}

\item {\em Complex}:~  SU$_L(N_f)\times$SU$_R(N_f) \to$ SU($N_f$). 

\item {\em Pseudo-real}:~ SU(2$N_f) \to $ Sp(2$N_f$).

\item {\em Real}:~ SU(2$N_f) \to $ SO(2$N_f$).

\end{itemize}

By Goldstone's theorem, the cosets of the above symmetry breaking
patterns determine the low-energy properties of these theories. Note that
precise details about the gauge group ${\cal G}$ do not enter at all. We are 
here getting a first bite of the universality that turns out to govern
the lowest-lying Dirac operator eigenvalues. 

As explained by Verbaarschot \cite{V1}, there exists a remarkable relation 
between the three symmetry breaking classes above and the classification
of chiral Random Matrix Theories ensembles. They are labelled by the so-called
Dyson indices $\beta$ as follows: The coset SU($N_f$)
of complex fermion representations corresponds to the 
{\em chiral Unitary Ensemble} chUE ($\beta=2$), the coset 
SU(2$N_f$)/Sp(2$N_f$) of pseudo-real representations corresponds to
the {\em chiral Orthogonal Ensemble} chOE ($\beta=1$), while the coset
SU(2$N_f$)/SO(2$N_f$) of real representations corresponds to the
{\em chiral Symplectic Ensemble} chSE of $\beta=4$. This exhausts the Dyson 
classification.\footnote{The labelling in terms of the integers $\beta$ is
related to a certain number occuring in the three different Random Matrix 
Theories; there is thus no ``missing'' class of $\beta=3$!}

These symmetry breaking patterns concern continuum fermions.
Staggered fermions, away from the continuum, have unusual patterns of chiral
symmetry breaking: all real and pseudo-real representations are precisely 
swapped. This has been known for a while for special cases
(see, $e.g.$ ref. \cite{HV1}), and has very recently
been shown to hold in general \cite{DHNS}. Another peculiarity
of staggered fermions is their apparent insensitivity to gauge field
topology, which we shall return to in detail below.

The relationship to Random Matrix Theory has caused some
confusion, although it can be stated very clearly. To get there,
the theory first has to be formulated for fixed topological charge
$\nu$. The order parameter for spontaneous chiral symmetry breaking
is formally defined by
$\Sigma \equiv \lim_{m\to 0}\lim_{V\to \infty} |\langle\bar{\psi}\psi
\rangle|$. To zoom in on the smallest
Dirac operator eigenvalues, and to see how the chiral condensate is formed,
one takes the chiral limit in an
unorthodox way \cite{LS}. 
First, the euclidean four-volume $V$ is kept finite, and it is
convenient to work at fixed ultraviolet cut-off $\Lambda$. 
We are thus temporarily working with bare
quantities which will have to be renormalized eventually. Both the
finite volume and UV cut-off are ideal set-ups
from the point of view of lattice gauge theory. 

Next, the quark masses
are chosen so that the pseudo-Goldstone bosons (``pions'')
are too light to fit inside the box, $i.e.~  m_{\pi} \ll 1/L$,
where $L$ is the linear extent of the 4-volume $V$. Now $V$ is
sent to infinity while the combination $m_i\Sigma V$ ($m_i$ being
quark masses) is kept finite -- 
it is always possible to satisfy these two constraints by taking 
$m_i$ small enough. Because $L$ is 
taken much larger than $1/\Lambda_{QCD}$,
the euclidean partition function is dominated by the pions. Higher-mass
states do
contribute to the partition function, but as the volume is sent to infinity
their contribution will eventually, for large enough volume, be
exponentially suppressed, of order $\exp[-ML]$, with $ML \gg 1$. In contrast,
the pions do not yield exponentially suppressed contributions:
their masses can be tuned to zero, and we precisely require
$m_{\pi}L \ll 1$ throughout. It is in this sense that the results
which will be reviewed below are exact: given any required degree of
accuracy, this accuracy can be achieved by simply tuning $V$ and $m_i$. 
The exact results hold in the limit.

Leutwyler and Smilga proposed to combine the above ingredients of
$V \gg 1/m_{\pi}^4$ and fixed topology
\cite{LS}. Because the euclidean partition function is dominated by
the pions, it follows from the coset of
chiral symmetry breaking. It is a non-renormalizable chiral Lagrangian
with, in principle, an infinite number of terms. However, because
of the peculiar finite-volume regime chosen, only the zero-momentum modes need
to be considered (again an approximation that can be made as accurate
as we wish by tuning $m_i$ and $V$) \cite{GL}. Let
us focus mainly on the symmetry breaking class relevant
for QCD. Then the effective partition function is dominated
by one single term in the effective Lagrangian:
\beq
Z ~=~ \int_{SU(N_f)}\!\!\! dU \exp\left[V\Sigma{\mbox{\rm ReTr}}
(e^{i\theta/N_f}{\cal M}U^{\dagger})\right]
\label{Zdef0}
\eeq
One crucial point here is the dependence on the vacuum angle $\theta$.
Because of the anomaly, the partition function does not depend 
on the quark mass matrix ${\cal M}$ and $\theta$ separately. A chiral
rotation (independent phase rotations of the left and right handed
fields) gives a linear shift in $\theta$, and since the partition
function is left unchanged by such a change of integration variables,
it can only depend on the invariant combination
$\exp[i\theta/N_f]{\cal M}$. We normally do not consider QCD 
with a non-zero vacuum angle, and here it is also just introduced at
a preliminary step, as a source of topological charge $\nu$, and as a
means for computing the partition function in sectors of fixed
$\nu$ \cite{LS}:
\beq
Z_{\nu} = \int_{U(N_f)}\!\!\!dU~ (\det U)^{\nu}\exp\!\left[V
\Sigma{\mbox{\rm ReTr}}({\cal M}U^{\dagger})\right] 
\label{Zdef}
\eeq
The projection onto fixed topological charge has been used as the
additional U(1)-factor that extends the integration over the zero-momentum
modes to U$(N_f)$. Note that $Z_{\nu}$ does not depend on 
$m_i, V$ and $\Sigma$ independently, but only on the combination 
\beq
\mu_i ~\equiv~ m_i V \Sigma ~.
\label{mu}
\eeq
We will see that this finite-volume scaling
is reflected in the Dirac operator spectrum. 

There are two distinct advantages for considering the theory at fixed
topological charge $\nu$. The first is that the zero-dimensional group
integral can be done analytically for all $N_f$ and
$\nu$ \cite{Brower}\footnote{An intermediate calculation by Berezin and
Karpelevich, from which 
this integral can be derived, dates back to 1958. 
I thank T. Wettig for informing me of this.}, 
\beq
Z_{\nu}(\{\mu_i\}) = 
\det[\mu_i^{j-1}I_{\nu+j-1}(\mu_i)]/\prod_{i>j}^{N_f}(\mu_i^2-\mu_j^2) 
\label{integral}
\eeq
where $I_n(x)$ is a modified Bessel function, and the matrix in the numerator
is of size $N_f\times N_f$.
The second is the big surprise: at fixed $\nu$
there is a connection to Random Matrix Theory.

\section{The relation to Random Matrix Theory}

As observed by Shuryak and Verbaarschot
\cite{ShV}, the group integral (\ref{Zdef}) can be rewritten as a Random Matrix
Theory partition function which has an 
uncanny resemblance to the original QCD path integral. Consider
\beq
\tilde{Z}_{\nu} \!\equiv\!
\int\!\! dW\! \prod_{f=1}^{N_{f}}{\det}\!(iM + m_f\!)\!
\exp\!\!\left[-\frac{N}{2} \mbox{Tr}\, V(M^2)\!\right] \label{RMTdef}
\eeq
where
\beq
M ~=~ \left( \begin{array}{cc}
              0 & W^{\dagger} \\
              W & 0
              \end{array}
      \right) ~.
\eeq
Here the integral is over complex matrices $W$ of rectangular
size $N\times(N+\nu)$. The $\tilde{m}_i$'s are dimensionless
numbers, and the potential $V(M^2)$ is unspecificied at this point. 
The matrix $M$ anticommutes with 
diag$\{1_{N},-1_{N+\nu}\}$, and as a consequence the eigenvalues of $M$ 
occur in pairs $\pm\tilde{\lambda}$ whenever $\tilde{\lambda}\neq 0$. 
Because of the
rectangular nature of $W$, the matrix $M$ also has precisely $\nu$
zero modes. The intuitive idea is the analogy with the determinant
of the Dirac operator for complex representation fermions: the matrix
$M$ has $\nu$ zero modes, it is chiral, and in  
(\ref{RMTdef}) one integrates over complex matrices.

But one is {\em not} simply trying to
replace the path integral over gauge potentials $A_{\mu}(x)$ by
zero-dimensional matrices. Instead, the precise relationship is as
follows. Take a ``microscopic limit'' of eq. (\ref{RMTdef}) in which
$\tilde{\mu}_i \equiv m_i(2N)\tilde{\rho}(0)$ is kept fixed as
$N \to \infty$. In that limit, and up to an irrelevant
($\mu_i$-independent) normalization,
\beq
Z_{\nu}[\{\mu_i\}] ~=~ \left.\tilde{Z}_{\nu}[\{\tilde{\mu}_i\}]
\right|_{\tilde{\mu}_{i}=\mu_i} ~.
\label{ZRMTeq}
\eeq
This was first demonstrated in ref. \cite{ShV}
for the case of a Gaussian potential in the $\beta=2$ chiral ensemble.
It follows from a series of universality theorems that
the identity holds for any generic choice of $V(M)$ \cite{ADMN}, and that
a fine tuning is required to reach other
universality classes that are of no obvious relevance here. Basically,
the domain of universality is determined by the condition that
$\tilde{\rho}(0) \neq 0$, as could have been expected from the manner
in which the microscopic limit is taken. Identities similar to
(\ref{ZRMTeq}) exist for the two other classes of chiral symmetry
breaking \cite{HV2}. As outlined above, 
the chiral Random Matrix Theory ensembles
are different, but there are 
analogous universality proofs for those cases \cite{SenerV}.

Having two partition functions coincide as in eq. (\ref{ZRMTeq}) may
not seem to give much information. But one should rather view
$Z_{\nu}[\{\mu_i\}]$ as a generating function for the chiral
condensate. Then the statement (\ref{ZRMTeq}) is highly non-trivial
since it means that in the infinite-volume limit where $\mu_i = m_i 
\Sigma V$ is kept fixed we can just as well compute the chiral condensate
from Random Matrix Theory. This observation
is also the basis for finally understanding why it is even possible
to compute the microscopic Dirac operator spectrum from Random Matrix Theory.
More about this below.

\section{The microscopic Dirac operator spectrum}

Although the two partition functions (\ref{Zdef}) and (\ref{RMTdef})
coincide in the microscopic limit, it is far from obvious that the
eigenvalues $\tilde{\lambda}$ of the matrix $M$ should be related to the 
eigenvalues of the Dirac operator $\lambda$. The eigenvalue spectrum of the 
Random Matrix Theory is {\em very} different from that of the Dirac
operator: for a Gaussian potential the eigenvalue density
of (\ref{RMTdef}) is the famous Wigner semi-circle, while the spectrum
of the Dirac operator is expected to approach $\rho(\lambda) \sim 
\lambda^3$ near the UV cut-off. One may search for {\em universality} 
that makes such
differences irrelevant. Here it means going to a scale near the origin
where both spectra, on that scale, are trivially identical (namely
constant, near $\lambda \sim 0$). This 
is possible, because by the Banks-Casher relation $\Sigma = \pi\rho(0)$
and the assumption of spontaneous chiral symmetry breaking we know that
$\rho(0) \neq 0$. Similarly, we must require of the potential in 
(\ref{RMTdef}) that it leads to $\tilde{\rho}(0)\neq 0$.
Then one can define microscopically rescaled variables, which for the
Dirac operator eigenvalues are $\zeta\equiv \lambda V\Sigma$ (cf.
eq. (\ref{mu})), and for Random Matrix Theory $\tilde{\zeta}\equiv
\tilde{\lambda}2N\pi\tilde{\rho}(0)$. A microscopic spectral density (and 
similarly for higher spectral correlators) of
the Dirac operator is analogously $\rho_s(\zeta) \equiv 
\rho(\zeta/(V\Sigma))/V$, which has a finite well-defined limit as
$V\to \infty$ \cite{ShV}. The microscopic density in the Random Matrix
Theory context is defined analogously, and it can be computed by various
technqiues. The first analytical expression, for $N_f$ massless
flavors in the $\beta=2$ universality class, was obtained in ref.
\cite{VZ}:
\beq
\rho_s(\zeta)\! =\! \frac{\zeta}{2}\!\left[J_{N_{f}+\nu}(\zeta)^2
\!-\!J_{N_{f}+\nu-1}(\zeta)J_{N_{f}+\nu+1}(\zeta)\right]
\label{rhofirst}
\eeq
which, as a first check, was found to reproduce the Leytwyler-Smilga
spectral sum rule for $e.g.$ $\langle\sum_n 1/\zeta_n^2\rangle$
\cite{VZ,Vsum}.
Corresponding expressions exist for $\beta=1,4$ \cite{V2}.

Quark masses can be included without difficulty by
performing a ``double-microscopic limit'' in which both $m_i$ and
eigenvalues $\lambda$ are blown up as in (\ref{mu}). 
The precise analytical form of the spectral correlators can then be 
worked out explicitly \cite{mass,mass1}.

There is a very compact formulation of all these results. A central object 
is the so-called kernel $K(\zeta_a,\zeta_b)$ from which all spectral 
correlation functions can be computed:
\beq
\rho_s(\zeta_1,\ldots,\zeta_k) = \det_{ab} K(\zeta_a,\zeta_b)
\eeq 
A ``Master Formula'' for this kernel is \cite{DAD},
\begin{eqnarray}
K(\zeta,\zeta') &=& 
C\sqrt{\zeta\zeta'}\prod_{f}^{N_{f}}
\sqrt{(\zeta^2+\mu_f^2)(\zeta'^2+\mu_f^2)}~\times \cr
&&\hspace{-1cm} Z_{\nu}^{(N_f+2)}(\{\mu\},i\zeta,i\zeta')/
Z_{\nu}^{(N_f)}(\{\mu\}) ~.
\end{eqnarray}
This formula encompasses all cases for $\beta=2$,
any number $N_f$ of fermions with masses $\mu_f$, and any $\nu$. Moreover,
although it is derived within Random Matrix Theory, we have used
the identity (\ref{ZRMTeq}) to express the r.h.s. entirely in terms
of field theory partition functions. There are closely related formulas
for the $\beta=4$ case \cite{DAD}.

One can also compute an infinite series of individual eigenvalue 
distributions, beginning with the smallest. 
They too can be expressed
in terms of the effective partition functions \cite{NDW}. 
The analytical formula is known in all generality: it gives the $k$'th
eigenvalue distribution for all three universality classes, for any number
of fermions $N_f$, and in an arbitrary sector of topological charge 
$\nu$.\footnote{The only exception is the $\beta=1$ class where, 
for technical reasons, $\nu$ must be odd in the general formula.}
To give a simple example, the quenched distribution of the first eigenvalue 
in a sector of arbitrary $\nu$ reads ($i,j=1,\ldots,\nu$) \cite{NDW}
\beq
P_{\mbox{\rm min}}(\zeta) = \frac{\zeta}{2}e^{-\zeta^2/4}
\det[I_{2+i-j}(\zeta)]
\eeq
for the $\beta=2$ universality class. 

One can perform
the sum over topological charge explicitly for all spectral correlation
functions, including the microscopic spectral density itself 
\cite{Dtop}. The result can for $N_f\!>\!0$ be written
in terms of the effective field theory partition functions
at vacuum angle $\theta = \pi$ when $k$ is odd:
\begin{eqnarray}
&&\!\!\!\!\!\!\!\!\!\!\!\!\rho_s(\zeta_1,\ldots,\zeta_k) = (-1)^{k[N_f/2]}
\prod_{j<l}^k 
(\zeta_j^2-\zeta_l^2)^2~\times \cr
&&\!\!\!\!\!\!\!\!\!\!\!\!\prod_{i=1}^k(|\zeta_i|\!\prod_f^{N_{f}}
(\zeta_i^2+\mu_i^2)) 
\frac{Z^{(N_{f}+2k)}
(\{\mu\},\!\{i\zeta_i\})_{\theta=k\pi}}{Z^{(N_{f})}(\{\mu\})_{\theta=0}}
\end{eqnarray}

\section{Derivation from chiral Lagrangians}

All microscopic spectral correlation functions, and all
probability distributions of individual eigenvalues, can thus be expressed 
in terms
of effective field theory partition functions. This puts universality
of these quantities on a very simple footing. It also gives a
strong hint that in fact Random Matrix Theory is not needed at all.
These results can indeed be derived entirely within 
the chiral Lagrangian framework. It hinges on the crucial identity
(\ref{ZRMTeq}), and an extension of it which we will discuss next.

One way to compute the spectral density of the Random Matrix Theory
(\ref{RMTdef}) is through the resolvent 
\beq
R_{\nu}(m_v) \equiv \left\langle{\mbox{\rm Tr}}\frac{1}{M-m_v}
\right\rangle ~,
\eeq
where $m_v$ is an external parameter. 
In the microscopic limit this 
is, in view of eq. (\ref{ZRMTeq}), nothing but the partially quenched
chiral condensate. To be precise, consider the supersymmetric method
for computing this quantity from the chiral Lagrangian \cite{BG}: One
adds a quark of mass $m_v+j$, and a bosonic quark of mass $m_v$. Then
the partially quenched chiral condensate is $\partial_j \left.
\ln Z_{\nu}\right|_{j=0}$. In a similar way,
one can modify the Random Matrix Theory (\ref{RMTdef}) and 
multiply the integrand by $\det[M+\tilde{m}_v+\tilde{j}]/\det[M+
\tilde{m}_v]$. Taking the microscopic limit, this 
leads to a supersymmetric generalization of the identity (\ref{ZRMTeq}),
where the chiral Lagrangian is the supersymmetric one associated
with partial quenching. Naively this Lagrangian would be based on the
Goldstone supermanifold $U(N_f+1|1)$, but in fact the proper choice
of integration domain is a more subtle issue, as discussed at length in
\cite{OTV,SS}. This way of proceeding
is only in order to establish why the Random Matrix Theory approach 
really does give exact answers for the microscopic Dirac operator
spectrum. If one is not interested in this, one can proceed completely
within the chiral Lagrangian framework, and simply compute the partially
quenched chiral condensate directly. 
As a simple $\beta=2$ example, consider the fully quenched case where, with
$\mu_v=\mu$, one finds \cite{OTV}:
\beq
\!\frac{\Sigma_{\nu}(\mu)}{\Sigma}\! =\! \mu\left(I_{\nu}(\mu)K_{\nu}(\mu)
\!+\! I_{\nu+1}(\mu)K_{\nu-1}(\mu)\right) + \frac{\nu}{\mu}
\label{IK}
\eeq
The microscopic spectral density of the Dirac operator is given by \cite{OTV}
\beq
\rho_s(\zeta) = \frac{1}{2\pi}{\mbox{\rm Disc}}
\left.\Sigma_{\nu}(\mu)\right|_{\mu=i\zeta}
\label{disc}
\eeq
which precisely agrees with (\ref{rhofirst})\footnote{The
microscopic spectral density by definition does not include the zero modes.}.

Similar expressions exist for higher spectral $k$-point functions.
See ref. \cite {TV} for the $\beta=2$ case. There is again exact
agreement with the Random Matrix Theory results.

\section{Replicas} 

It is interesting that one can get quite far by
an alternative formulation of partial quenching based on the
replica method. The fermion determinant is then removed by 
analytic continuation in the number of flavors, so that eventually
one can take the limit $N_v\to 0$. Here $N_v$ denotes a number
of additional unphysical flavors added to the theory with $N_f$ physical 
quarks. 

The first observation is that now the identity (\ref{ZRMTeq}) 
as it stands is all that
is needed in order to show equivalence to the Random Matrix Theory
approach. The integration in eq. (\ref{Zdef}) is over U($N_f+N_v$),
and the limit $N_v \to 0$ is taken after differentiating
w.r.t. $\mu_v$. Small-mass and large-mass expansions can be performed
in this way \cite{DS,DV}. 
As an example, for the quenched chiral condensate
the replica method yields precisely small-mass and large-mass 
series expansions of the expression (\ref{IK}), whose inversion in turns
yields the same microscopic spectral density as was originally derived
from Random Matrix Theory. It is a highly non-trivial check on the
consistency of the whole formalism that these results all coincide.

\section{Lattice results}

This review has not proceeded in historical order. The
derivations of the microscopic Dirac operator spectrum based 
directly on the partially quenched chiral Lagrangian came after
the first lattice study \cite{Bbetal} already had
demonstrated the viability of the Random Matrix Theory approach.
Since then there have been numerous lattice comparisons
with the exact analytical expressions, for different gauge groups,
different representation quarks in both quenched and full theories
\cite{moreMC,Lang,DHNS}.
It is impossible to do justice to all that work here, but a few
points can be highlighted.

\begin{figure}
\includegraphics[scale=0.29]{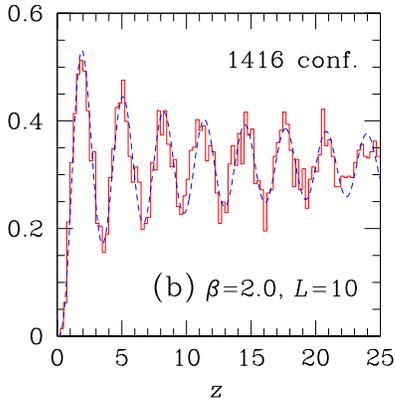}
\caption{The first Monte Carlo measurement of the microscopic spectral
density of the Dirac operator \cite{Bbetal}, for gauge group SU(2) and 
quenched staggered fermions. The dashed curve is the analytical prediction.}
\end{figure}

A quite systematic study of the smallest eigenvalue distributions in
all three universality classes and for zero and non-zero
topological charge showed perfect agreement with the analytical
predictions \cite{nu}. The distribututions of a whole sequence of smallest 
eigenvalues have been compared with the analytical formulas in ref.
\cite{indiv}. One sees
clearly how the microscopic spectral density is built up of the individual
distributions. 

Also higher-point spectral correlators have been checked
\cite{Bbetal,Gockeler}, again
with spectacularly good agreement with the analytical predictions.
Extensions to the regime of chiral perturbation theory
have also been made \cite{chiral}.

Lattice calculations with staggered fermions compare only with the 
analytical expressions for the sector with $\nu=0$. This makes sense,
since it has been known for long \cite{SVink} that staggered fermions
do not have exact zero modes at usual gauge couplings. Even
if one selects distinct topologically non-trivial gauge field sectors,
the microscopic Dirac spectrum seems completely
unaffected, and always agrees with the $\nu=0$ predictions \cite{DHNR}.
There is a much more direct way of understanding this. The symmetry
breaking patterns of staggered fermions away from the continuum always
have an additional U(1) factor\footnote{They are U($N)\times$U($N)
\to$ U($N$) [complex], U($2N)\to$ Sp($2N$) [real], and   
U($2N)\to$ SO($2N$) [pseudo-real]} which
for the zero momentum modes is completely equivalent to a projection
on the $\nu=0$ sector \cite{D01,DHNS}. So we {\em should} only compare
with what can be viewed as $\nu=0$ predictions of the theory without the
extra U(1) factor. If one sees sensitivity to 
topology \cite{FHL}, then continuum
flavor symmetries are beginning to be recovered.

Instead of
looking directly at the smallest Dirac operator eigenvalues, one can also
consider
derived quantities such as the (mass-dependent) chiral condensate
\cite{V3,DEHN,Pilar,Tom,Bern}
\beq
\frac{\Sigma_{\nu}(\mu)}{\Sigma} = 2\mu\int_0^{\infty}d\zeta 
\frac{\rho_s^{(\nu)}(\zeta,\mu)}{\zeta^2+\mu^2} + \frac{\nu}{\mu}~,
\label{Sigmaformula}
\eeq
or higher chiral susceptibilities, such as \cite{DEHN}
\beq
\frac{\omega_{\nu}(\mu)}{V\Sigma^2} = 4\mu^2\int_0^{\infty}\!d\zeta
\frac{\rho_s^{(\nu)}(\zeta,\mu)}{(\zeta^2+\mu^2)^2} + \frac{2\nu}{\mu^2}
\eeq
In the quenched and partially quenched cases 
there is
a one-to-one correspondence between $\Sigma(\mu)_{\nu}$ and the microscopic
spectral density $\rho_s^{(\nu)}(\zeta)$. But: 
$\Sigma_{\nu}(\mu)$ has a short-distance singularity that is not taken
into account in the above description. At fixed UV cut-off
$\Lambda$ this is not a problem when one takes the infinite-volume limit,
as one should in order to compare with (\ref{IK}). Corrections are of
the form $Am\Lambda^2 + Bm^3\ln\Lambda$, where $A$ and $B$ are constants.
This looks scary, but we are taking the limit in which $\mu=m\Sigma V$
is kept fixed as $V\to\infty$. So these UV corrections actually vanish;
they are suppressed by inverse factors of the volume, and really read
$(A/\Sigma)(\mu/V)\Lambda^2 + (B/\Sigma^3)(\mu^3/V^3)\ln\Lambda$. If the 
volume $V$ is not large enough these terms can be annoying,
and then one should subtract them. 

There are cases where very, very few Dirac operator eigenvalues
play any r\^{o}le in building up the chiral condensate (\ref{IK}) in the
interesting mass range. As an example,
let us compare the exact result (\ref{IK})
with what one gets by keeping only {\em one single Dirac operator eigenvalue}
in the integral of eq. (\ref{Sigmaformula}) for $\nu=0$. We then simply 
replace $\rho_s^{(0)}(\zeta)$ in the integrand of (\ref{Sigmaformula})
by the probability distribution of just one single Dirac operator
eigenvalue, the smallest, $
P_{\mbox{\rm min}}^{(0)}(\zeta) = (\zeta/2)e^{-\zeta^2/2}$.
This may seem like an absurd truncation, and of course an unnecessary one,
since we know the full analytical form of $\rho_s^{(0)}(\zeta)$. But the
comparison between this approximation, and the exact answer is shown
in figure 2. 

\begin{figure}
\includegraphics[width=7.5cm]{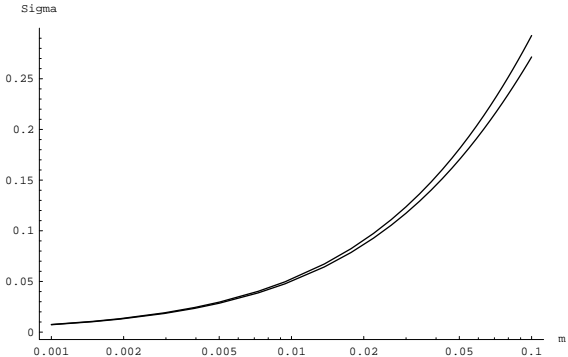}
\caption{The exact quenched chiral condensate (here $m$ stands for $\mu$)
for $\nu=0$, compared with keeping only one single eigenvalue
in eq. (\ref{Sigmaformula}).}
\end{figure}

The two curves are basically indistinguishable (up to
$\mu \sim 0.1$ the difference is less than 5\%, 
decreasing fast as $\mu$ is lowered). If one sees
good scaling of the $\nu=0$ quenched chiral condensate
in such a mass range, then one is probing, and having good
statistics of, only one single Dirac operator eigenvalue.  
Of course, we could just as well have included
all the other eigenvalues, and then the relation would be exact.
For extracting
$\Sigma$ in this way it seems more logical to measure the distributions
of the Dirac operator eigenvalues directly. When $\nu\neq 0$ the 
eigenvalues are shifted further away from the origin, 
the integral 
(\ref{Sigmaformula}) is not as infrared sensitive, and the brutal
approximation of keeping only the first eigenvalue is not nearly as good. 
The error is easily assessed if we rewrite, for $\nu \neq 0$ \cite{DEHN},
\beq
\frac{\Sigma_{\nu}(\mu)}{\Sigma} = \frac{\nu}{\mu} + 2\mu\left\langle
\sum_{n>0}\frac{1}{\zeta_n^2}\right\rangle + \ldots~,
\eeq
which shows that to leading order one is simply trying to measure the first
Leutwyler-Smilga sum rule \cite{LS}, $\langle\sum_n 1/\zeta_n^2\rangle
= 1/(4\nu)$. It
seems advantageous to instead measure the first few eigenvalue
distributions, or, for lower statistics data, at least individual averages. 
That way one avoids contaminating the result with
the larger eigenvalues which at the limited finite volume will have
distributions that are incompatible with the analytical formulae.

\section{Beyond Random Matrix Theory}

We have emphasized that the results reviewed here are exact
in the sense that they can be made as accurate as we wish by
tuning the quark masses $m_i$ and the volume $V$. It immediately suggests that 
it is possible to compute 
corrections to this scaling, which would
improve convergence and bring in more terms from the chiral Lagrangian. 
This is indeed the case. A computational
framework was laid out by Gasser and Leutwyler, who called
it the $\epsilon$-expansion 
\cite{GL}.\footnote{The quantity $\epsilon$ counts orders in $1/L$,
where $L$ is the linear extent of the volume. There is no relation to
the expansion around critical dimensions.}

The idea is simple. Instead of keeping only the zero-momentum mode
in the chiral Lagrangian, one includes the modes of non-zero momentum
in a perturbative manner. Thus, one starts with the full chiral
Lagrangian and then separates out the zero-momentum modes of $U(x)$:
\beq
U(x) ~=~ ue^{i\sqrt{2}\xi(x)/F}u
\label{Udef}
\eeq
where $u \in$ SU($N_f$) is a space-time independent collective
field, and $\xi(x)$ contains only modes of non-zero momentum. To leading
order, where one simply ignores $\xi(x)$, this gives the effective
partition function (\ref{Zdef0}) after identifying $U \equiv u^2$.
To next order the kinetic energy term contributes, and one resorts
to the usual loop expansion in the fluctuation field $\xi(x)$. The
modes of zero momentum are still treated {\em exactly}. So this
expansion is a combination of the exact, non-perturbative, 
leading-order result for $V \to \infty$, and a perturbative expansion
where each term should be suppressed by powers of $1/L \equiv 1/V^{1/4}$.
In 4 dimensions the expansion is even better behaved, as the first
correction goes as $1/L^2$, rather than just $1/L$ \cite{GL,N}.
Leading terms are what we alternatively could obtain from Random
Matrix Theory. The corrections take us beyond.
 
The one-loop correction to the chiral condensate computed in this
way can be re-absorbed into a volume-dependent $\Sigma_{eff}(V)$ \cite{GL},  
\beq
\frac{\Sigma_{eff}(V)}{\Sigma} = 1 + \frac{N_f^2-1}{N_f}\frac{1}{F^2}
\frac{\beta_1(L_i/L)}{L^2} 
\label{Sigmashift0}
\eeq
where $\beta_1$ depends on the geometry \cite{N,HL}. This carries directly
over to the Dirac operator eigenvalues, on account of the relation
(\ref{disc}). From the small $1/L^2$-correction one can thus, with
sufficient statistics, also measure the pion decay constant $F$ from
the eigenvalue distributions.

The quenched $\epsilon$-expansion meets the usual difficulties
of the quenched theory. We outline the problem 
here using the replica version of quenched chiral perturbation theory
\cite{DS1} because it is simpler to explain. 
Then (\ref{Udef}) still defines the split between modes of zero and
non-zero momentum, but now $U(x)\in $U($N_v)$. Let $\Xi(x) \equiv
{\mbox{\rm Tr}}\xi(x)$
denote the flavor singlet fluctuations. The lowest-order chiral Lagrangian
in a sector of fixed topology reads \cite{D01}
\begin{eqnarray}
&&\!\!\!\!\!\!\!\!\!\!\!\!\!{\mbox{\rm Tr}}\!\left[\frac{1}{2}
\partial_{\mu}\xi(x)\partial^{\mu}\xi(x)\!+\! 
\frac{m_0^2}{6}\Xi(x)^2\!+\! 
\frac{\alpha}{6}\partial_{\mu}\Xi(x)\partial^{\mu}\Xi(x)\right.
\cr
&&\!\!\!\!\!\!\!\!\!\!\!\! -\frac{\Sigma}{2}\left.{\cal M}(U + 
U^{\dagger})\left(V - 
\frac{1}{F^2}\int\!dx~\xi(x)^2\right)\right] ~.
\end{eqnarray}
and the propagator of $\Xi(x)$ is thus modified due to the $m_0$-term
(the precise form is given in \cite{DS1}), and one again takes the limit
$N_v\to 0$ after having differentiated with respect to the sources. 
A new scale is introduced
by the $m_0$-term, and in order to proceed perturbatively one must assume
that $m_0^2/(4\pi F^2)$ is small. At one-loop level the quenched analog
of the effective volume-dependent $\Sigma_{eff}(V)$ is \cite{D01}
\beq
\frac{\Sigma_{eff}(V)}{\Sigma}\!=\! 
1\!-\! \frac{1}{3F^2}\!\left[\frac{\alpha\beta_1}{L^2}
- \frac{m_0^2}{8\pi^2}\ln(L)\!\right]
\eeq
In the full theory logarithms are typically multiplied by $m_{\pi}^2$,
and hence in this regime suppressed by at least $1/L^2$. Here it survives
at fixed $m_0$, indicating difficulties with the whole expansion. It is
not a quenched chiral logarithm, but, 
in this regime, a quenched finite-volume logarithm
that signals the limitations of the expansion.

The $\epsilon$-expansion has also been worked out for
correlation functions of the full theory \cite{N,H}, and very recently
these results have been generalized to the quenched
case \cite{DDHJ}. By measuring correlation functions in this regime
one can extract the low-energy constants of QCD.

\section{Conclusions}

The microscopic tail of the Dirac operator spectrum is computable by
a precise comparison with the associated effective field theory. The
result is a series of exact expressions for observables in finite-volume
gauge theories. This is a highly unusual situation for such strongly
coupled theories, and one can make good use of these results. In
particular:

$\bullet$ Reliability of fermion algorithms can be checked by comparing with
the exact expressions.

$\bullet$ Gauge field topology gives distinct predictions, and is probed at 
the ``quantum level''.

$\bullet$ Physical observables derivable from the microscopic spectral
correlation functions are known in exact analytical forms. Simple models,
ans\"{a}tze, or unknown extrapolations for the small-mass finite-volume
behavior can be replaced by the correct analytical
expressions.

$\bullet$ By going to {\em unphysical regimes} we can extract {\em physical
observables}. This includes unphysical volumes, and, if need be, fixed
gauge field topology. The most extreme example was given: From the
distribution of just one single Dirac eigenvalue we learn {\bf i}) whether
spontaneous chiral symmetry occurs in the infinite-volume theory, {\bf ii})  
which symmetry breaking class it belongs to, {\bf iii}) the value of
the infinite-volume chiral condensate $\Sigma$, and finally, by looking
even closer, {\bf iv}) the pion decay constant $F$. If one is willing
to push it, in principle the infinite series of parameters of the
chiral Lagrangian can be probed by the Dirac operator
eigenvalues.


\begin{thebibliography}{9}

\bibitem{V1}
J.~Verbaarschot,
Phys.\ Rev.\ Lett.\  {\bf 72} (1994) 2531.

\bibitem{HV1}
M.~A.~Halasz and J.~J.~Verbaarschot,
Phys.\ Rev.\ Lett.\  {\bf 74} (1995) 3920.

\bibitem{DHNS}
P.~H.~Damgaard {\em et al.},
hep-lat/0110028.


\bibitem{LS}
H.~Leutwyler and A.~Smilga,
Phys.\ Rev.\ D {\bf 46} (1992) 5607.

\bibitem{GL}
J.~Gasser and H.~Leutwyler,
Phys.\ Lett.\ B {\bf 188} (1987) 477.

\bibitem{Brower}
R.~Brower {\em et al.},
Nucl.\ Phys.\ B {\bf 190} (1981) 699,
A.~D.~Jackson {\em et al.},
Phys.\ Lett.\ B {\bf 387} (1996) 355.

\bibitem{ShV}
E.~V.~Shuryak and J.~J.~Verbaarschot,
Nucl.\ Phys.\ A {\bf 560} (1993) 306.

\bibitem{ADMN}G. Akemann {\em et al.}
Nucl.\ Phys.\ B {\bf 487} (1997) 721;
{\em ibid.} \ B {\bf 519} (1998) 682

\bibitem{SmV}
A.~Smilga and J.~J.~Verbaarschot,
Phys.\ Rev.\ D {\bf 51} (1995) 829.

\bibitem{HV2}
M.~A.~Halasz and J.~J.~Verbaarschot,
Phys.\ Rev.\ D {\bf 52} (1995) 2563.

\bibitem{SenerV}
M.~K.~Sener and J.~J.~Verbaarschot,
Phys.\ Rev.\ Lett.\  {\bf 81} (1998) 248.


\bibitem{DAD}
P.~H.~Damgaard,
Phys.\ Lett.\ B {\bf 424} (1998) 322
G.~Akemann and P.~H.~Damgaard,
Nucl.\ Phys.\ B {\bf 528} (1998) 411;
Phys.\ Lett.\ B {\bf 432} (1998) 390.

\bibitem{VZ}
J.~J.~Verbaarschot and I.~Zahed,
Phys.\ Rev.\ Lett.\  {\bf 70} (1993) 3852.

\bibitem{Vsum}
J.~Verbaarschot,
Phys.\ Lett.\ B {\bf 329} (1994) 351.


\bibitem{V2}
J.~Verbaarschot,
Nucl.\ Phys.\ B {\bf 426} (1994) 559,
T.~Nagao and P.~J.~Forrester,
Nucl.\ Phys.\ B {\bf 435} (1995) 401.


\bibitem{mass}
J.~Jurkiewicz {\em et al.},
Nucl.\ Phys.\ B {\bf 478} (1996) 605,
P.~H.~Damgaard and S.~M.~Nishigaki,
Nucl.\ Phys.\ B {\bf 518} (1998) 495,
T.~Wilke {\em et al.},
Phys.\ Rev.\ D {\bf 57} (1998) 6486.

\bibitem{mass1}
T.~Nagao and S.~M.~Nishigaki,
Phys.\ Rev.\ D {\bf 62} (2000) 065006;
{\em ibid.} {\bf 62} (2000) 065007,
G.~Akemann and E.~Kanzieper,
Phys.\ Rev.\ Lett.\  {\bf 85} (2000) 1174,
F.~Abild-Pedersen and G.~Vernizzi,
hep-th/0104028.



\bibitem{NDW}S.~M.~Nishigaki {\em et al.},
Phys.\ Rev.\ D {\bf 58} (1998) 087704,
P.~H.~Damgaard and S.~M.~Nishigaki,
Phys.\ Rev.\ D {\bf 63} (2001) 045012.


\bibitem{Dtop}
P.~H.~Damgaard,
Nucl.\ Phys.\ B {\bf 556} (1999) 327,
G.~Akemann and P.~H.~Damgaard,
Nucl.\ Phys.\ B {\bf 576} (2000) 597.


\bibitem{BG}
C.~W.~Bernard and M.~F.~Golterman,
Phys.\ Rev.\ D {\bf 49} (1994) 486.

\bibitem{OTV}
J.~C.~Osborn {\em et al.},
Nucl.\ Phys.\ B {\bf 540} (1999) 317,
P.~H.~Damgaard {\em et al.},
Nucl.\ Phys.\ B {\bf 547} (1999) 305,
D.~Toublan and J.~J.~Verbaarschot,
Nucl.\ Phys.\ B {\bf 560} (1999) 259.


\bibitem{TV}
D.~Toublan and J.~J.~Verbaarschot,
Nucl.\ Phys.\ B {\bf 603} (2001) 343.

\bibitem{SS}
S.~R.~Sharpe and N.~Shoresh,
hep-lat/0108003.

\bibitem{DS}
P.~H.~Damgaard and K.~Splittorff,
Nucl.\ Phys.\ B {\bf 572} (2000) 478,
P.~H.~Damgaard,
Phys.\ Lett.\ B {\bf 476} (2000) 465.

\bibitem{DV}
D.~Dalmazi and J.~J.~Verbaarschot,
Nucl.\ Phys.\ B {\bf 592} (2001) 419.

\bibitem{Bbetal}
M.~E.~Berbenni-Bitsch {\em et al.},
Phys.\ Rev.\ Lett.\  {\bf 80} (1998) 1146.

\bibitem{moreMC}
M.~E.~Berbenni-Bitsch {\em et al.},
Phys.\ Rev.\ D {\bf 58} (1998) 071502,
P.~H.~Damgaard {\em et al.},
Phys.\ Lett.\ B {\bf 445} (1999) 366,
R.~G.~Edwards {\em et al.},
Phys.\ Rev.\ D {\bf 60} (1999) 077502,

\bibitem{Lang}
F.~Farchioni {\em et al.},
Nucl.\ Phys.\ B {\bf 549} (1999) 364,
B.~A.~Berg {\em et al.},
Phys.\ Rev.\ D {\bf 63} (2001) 014504,
B.~A.~Berg {\em et al.},
Phys.\ Lett.\ B {\bf 514} (2001) 97.

\bibitem{nu}
R.~G.~Edwards {\em et al.},
Phys.\ Rev.\ Lett.\  {\bf 82}, 4188 (1999).

\bibitem{indiv}
P.~H.~Damgaard {\em et al.}
Phys.\ Lett.\ B {\bf 495} (2000) 263.

\bibitem{Gockeler}
J.~Z.~Ma {\em et al.},
Eur.\ Phys.\ J.\ A {\bf 2} (1998) 87,
M.~Gockeler {\em et al.},
Phys.\ Rev.\ D {\bf 59} (1999) 094503.

\bibitem{chiral}
M.~E.~Berbenni-Bitsch {\em et al.}
Phys.\ Lett.\ B {\bf 466} (1999) 293,
M.~Gockeler {\em et al.},
hep-lat/0105011.


\bibitem{SVink}
J.~Smit and J.~C.~Vink,
Nucl.\ Phys.\ B {\bf 286} (1987) 485.


\bibitem{DHNR}
P.~H.~Damgaard {\em et al.},
Phys.\ Rev.\ D {\bf 61} (2000) 014501.

\bibitem{D01}
P.~H.~Damgaard,
Nucl.\ Phys.\ B {\bf 608} (2001) 162.

\bibitem{FHL}
F.~Farchioni {\em et al.},
Phys.\ Lett.\ B {\bf 471} (1999) 58



\bibitem{V3}
J.~J.~Verbaarschot,
Phys.\ Lett.\ B {\bf 368} (1996) 137.

\bibitem{DEHN}
P.~H.~Damgaard {\em et al.},
Phys.\ Rev.\ D {\bf 61} (2000) 094503.

\bibitem{Pilar}
P.~Hernandez {\em et al.},
Phys.\ Lett.\ B {\bf 469} (1999) 198.

\bibitem{Tom}
T.~DeGrand,
Phys.\ Rev.\ D {\bf 63} (2001) 034503.

\bibitem{Bern}
P.~Hasenfratz {\em et al.},
hep-lat/0109007.


\bibitem{N}
H.~Neuberger,
Nucl.\ Phys.\ B {\bf 300} (1988) 180.


\bibitem{DS1}
P.~H.~Damgaard and K.~Splittorff,
Phys.\ Rev.\ D {\bf 62} (2000) 054509.

\bibitem{H}
F.~C.~Hansen,
Nucl.\ Phys.\ B {\bf 345} (1990) 685.

\bibitem{HL}
P.~Hasenfratz and H.~Leutwyler,
Nucl.\ Phys.\ B {\bf 343} (1990) 241.

\bibitem{DDHJ}
P.~H.~Damgaard {\em et al.}
hep-lat/0110170.

\end{thebibliography}
\end{document}